\documentclass[amsmath,amssymb,
               aps,
               prl,
               twocolumn,
               superscriptaddress,
               showpacs,
               floatfix,
               longbibliography,
               nofootinbib,
               preprintnumbers
]{revtex4-2}

\usepackage[utf8]{inputenc}
\usepackage{graphicx}
\usepackage{color}
\usepackage{dcolumn}
\usepackage{bm}
\usepackage[english]{babel}
\usepackage{lipsum}
\usepackage[dvipsnames]{xcolor}
\usepackage{braket}
\usepackage{siunitx}
\usepackage{mathtools}
\usepackage{multirow}
\usepackage{verbatim}
\usepackage[colorlinks=true,
            linkcolor=blue,
            citecolor=blue,
            urlcolor=blue]{hyperref}
\usepackage{comment}
\usepackage{xspace}

\newcommand{\emax}{\text{e}_\text{1max}\xspace}
\newcommand{\etmax}{\text{e}_\text{3max}\xspace}

\newcommand{\magic}{1.8/2.0 (EM)\xspace}
\newcommand{\go}{{\Delta}{\rm NNLO}_{\rm GO}(394)}

\newcommand{\ai}{\emph{ab initio}}

\begin{document}

\title{The iconic $^{238}$U: \emph{ab initio} nuclear structure theory towards the limit\\ of the 
periodic table}%

\author{Alberto~Scalesi}
    \email{alberto.scalesi@chalmers.se}
    \affiliation{Department of Physics and Astronomy, Chalmers University of Technology, SE-412 96 G\"oteborg, Sweden}
\author{Thomas~Duguet}
	\affiliation{Universit\'e Paris-Saclay, CEA, IRFU, 91191 Gif-sur-Yvette, France}
\author{Vittorio~Som\`a}
	\affiliation{Universit\'e Paris-Saclay, CEA, IRFU, 91191 Gif-sur-Yvette, France}

\date{\today}

\begin{abstract}
The \textit{ab initio} description of heavy and superheavy nuclei constitutes one of the holy grails of nuclear theory, bearing on the synthesis of the heaviest elements and the limits of nuclear stability. Over the last fifteen years, many-body expansion methods, whose numerical cost scales polynomially with system size, have extended first-principles calculations to medium-mass nuclei and a few spherical closed-shell heavy systems. The largest portion of the nuclear chart is however composed of heavy deformed doubly open-shell nuclei and has remained completely out of reach. This is due to two major obstacles: (i) the huge computational cost of beyond mean-field calculations in very large single-particle bases, and (ii) a dubious collapse of the mean-field energy at large prolate deformation. While a highly efficient numerical implementation of the novel deformed self-consistent Green's function formalism removes the first difficulty, the second is cured by the inclusion of many-body correlations beyond the deformed mean field.
Presenting the first \textit{ab initio} calculation of the iconic $^{238}$U nucleus, this work brings the upper-end of the nuclear chart within reach of theoretical predictions based on first principles.
\end{abstract}

\maketitle

\paragraph*{Introduction.}
The origin of the heaviest elements in the universe is one of the most fascinating open questions in modern science. Approximately half of all the elements heavier than iron are expected to be forged through the rapid neutron-capture process ($r$-process)~\cite{Cowan21}, whose astrophysical site was recently confirmed by the observation of a kilonova following the neutron star merger GW170817~\cite{LIGOScientific:2017vwq,Kasen:2017sxr}. During the $r$-process, very neutron-rich nuclei are produced via rapid neutron captures until undergoing a $\beta$-decay to produce the element with one more proton, driving matter into a vast landscape of neutron-rich, largely unexplored isotopes far from the valley of stability. The final abundance pattern of heavy elements --- including gold, platinum, and uranium --- is exquisitely sensitive to nuclear properties such as masses, $\beta$-decay rates, neutron-capture cross sections, and fission half-lives along the $r$-process path~\cite{MUMPOWER201686}.
Many of the nuclei involved display a large intrinsic deformation directly affecting level densities, collective excitations, and fission fragment distributions, all of which feed into astrophysical network calculations~\cite{Arnould:2007gh}. In the actinide and superheavy region in particular, fission recycling --- where heavy nuclei fission and their fragments are recaptured --- plays a decisive role in shaping the final elemental abundances~\cite{Wang:2020qre}. Yet the nuclear structure input for these calculations relies almost entirely on phenomenological models~\cite{Moller16,Goriely:2009zzb}, whose predictive power far from stability is difficult to assess.

Beyond astrophysics, superheavy nuclei represent a frontier of nuclear science in their own right. While the latest element having entered the periodic table in 2016 possesses $Z=118$ protons (Oganesson), a predicted island of stability around $Z \sim 114$--$126$~\cite{OhrstromReedijk,Oganessian_2015} --- where shell effects are expected to stabilize otherwise short-lived nuclei against fission --- is motivating intensive experimental searches at facilities worldwide~\cite{Hofmann:2016tix}. The interplay of shell structure and intrinsic deformation in this region determines which isotopes can exist, how long they live, and what their decay properties are. Ground-state deformations in the actinide and transactinide region are large~\cite{Moller16}, with nuclei such as $^{238}$U exhibiting a pronounced intrinsic prolate shape that strongly influences its spectroscopy and reaction dynamics. New experimental facilities such as FRIB~\cite{Balantekin14}, SPIRAL2~\cite{Gales:2010dsu} and FAIR~\cite{Spiller:2006gj} are set to produce and probe increasingly exotic heavy isotopes. 
Complementarily, ultra-relativistic heavy-ion collisions at RHIC and the LHC are now also able to probe inter-nucleon correlations underlying nuclear intrinsic deformations~\cite{Giacalone2018a,STAR2024a,Giacalone2025a,Duguet:2025hwi,Bofos:2026huw}. All this is creating an urgent need for theoretical predictions grounded in the fundamental theory of the strong force, quantum chromodynamics (QCD). 

First-principles \textit{ab initio} calculations beyond the lightest nuclei build on (i) two- and three-nucleon interactions rooted into QCD through chiral effective field theory ($\chi$EFT)~\cite{Epelbaum09,Machleidt11} and (ii) polynomial-scaling many-body methods used to solve $A$-body Schr\"odinger's equation with sub-percent accuracy~\cite{Hergert20}. In the last 15 years, the development of expansion methods has dramatically increased the reach of \textit{ab initio} calculations over the mass table, taking them from Carbon ($Z=6$) all the way to  Zirconium ($Z=40$) for deformed doubly open-shell nuclei~\cite{Hu24}, to Tin ($Z=50$) for spherical singly open-shell nuclei~\cite{Demol26,Vernik2026} and to Lead ($Z=82$) for spherical doubly closed-shell nuclei~\cite{Hu22,Bonaiti2025heavy}. In spite of this extraordinary progress, the vast majority of nuclei are heavy deformed doubly open-shell systems that have remained completely out of reach so far. The present work takes a decisive step towards the upper-end of the nuclear chart by taking on the two challenges that have so far prevented controlled {\it ab initio} calculations of (very) heavy deformed nuclei, i.e.\ (i) the enormous numerical challenge associated with the necessity to use very large bases to produce converged many-body calculations and (ii) a dubious collapse of the mean-field energy at large prolate deformations.  The proof of principle presented here is based on the first-ever \emph{ab initio} calculation of the iconic $^{238}$U nucleus.

\paragraph*{Methodology.}
Given the nuclear Hamiltonian $H$, the present work relies on the development of a novel expansion method  to solve $A$-body Schr\"odinger's equation 
\begin{equation}
H | \Psi_{k} \rangle = E_k | \Psi_{k} \rangle \, , \label{Sch}
\end{equation}
coined as {\it deformed Self-consistent Green's function} (dSCGF) theory~\cite{PRC}. 
This approach generalizes its existing spherical counterparts restricted to doubly closed-shell~\cite{Raimondi18} and singly open-shell~\cite{Soma11,Soma14a} nuclei. 
The dSCGF($n$) method is implemented at orders $n=1,2,3$ in the algebraic diagrammatic construction scheme~\cite{Schirmer83}, the latter two delivering solutions at a few-percent and sub-percent accuracy, respectively. 
Present {\it ab initio} calculations employ representative $\chi$EFT nuclear Hamiltonians containing  two- and three-nucleon interactions, namely the \magic{}~\cite{Hebeler11a} and the $\go{}$~\cite{Jiang20}.

Paralleling the recent development of deformed coupled cluster theory~\cite{Novario20}, the multi-reference in-medium similarity renormalization group approach~\cite{Yao20} and the projected-generator-coordinate-method-based perturbation theory~\cite{Frosini22a,Frosini22b,Frosini22c} to address doubly open-shell nuclei, the highly efficient numerical optimization of the presently developed dSCGF method overcomes computational bottlenecks to make it possible, for the first time, to converge calculations of heavy deformed nuclei with respect to the employed basis dimension~\cite{PRC}. Indeed, \textit{ab initio} calculations rely on representing a given $k$-nucleon (kN) operator, e.g.\ the three-nucleon interaction, on a truncated basis of the $k$-body Hilbert space. This is done starting from a one-body spherical harmonic oscillator basis (sHO) characterized by an optimal frequency  $\hbar\omega$ and truncated to a given finite dimension characterized by the parameter $\emax$. Typically, 3N interaction matrix elements must be further limited to three-body basis states characterized by $\etmax \leq 3\,\emax$. Assessing the convergence of  physical observables with respect to both $\emax$ and $\etmax$ constitutes a key aspect of any \textit{ab initio} nuclear structure calculation and represents a major challenge in (very) heavy deformed nuclei. This is especially true when performing highly accurate many-body, e.g.\ dSCGF($3$), calculations~\cite{PRC}. Furthermore, the 3N interaction need to be approximated as an effective 2N interaction via a rank-reduction (RR) method~\cite{Frosini21} whose impact must also be controlled. 

The dSCGF calculation of $^{238}$U builds a fully correlated solution of Eq.~\eqref{Sch} on top of a Slater determinant solution of the {\it deformed} Hartree-Fock (dHF) mean-field equation carrying an intrinsic axial quadrupole deformation $\beta^{\text{dHF}}_2$. This dHF state actually delivers the dSCGF($1$) approximation to the full dSCGF expansion. By breaking rotational invariance of the Hamiltonian, dHF efficiently captures strong, so-called static, angular correlations among the nucleons that strongly imprint the two-body correlation function of the full solution $| \Psi_{k} \rangle$~\cite{Duguet:2025hwi,Bofos:2026huw,Bofos:2026nmg}. Building on this dHF starting point, dSCGF($2,3$) solutions themselves carry a non-zero intrinsic quadrupole deformation $\beta_2^\text{dSCGF(2,3)}$. As such, each dSCGF solution describes an {\it intrinsic} state of $^{238}$U, i.e.\ a linear combination of a specific sequence of $J^{\pi}=0^+, 2^+, 4^+\ldots$ eigenstates making up a rotational band. While each member of that band can eventually be resolved via angular momentum projection~\footnote{While this has been formulated~\cite{Duguet15a,Qiu17,Qiu18} and implemented~\cite{Hagen22} for other expansion methods, it remains to be done for dSCGF theory.}, the intrinsic state already delivers an excellent account of bulk properties of the $J^{\pi}=0^+$ bandhead~\cite{Frosini22b,Frosini22c,Hagen22,Hu:2024jih,Sun25,Bofos:2026nmg}.

Following this procedure, one can obtain several dSCGF($n$) solutions describing a subset of the quantum states of the nucleus, i.e.\ the ground-state and so-called shape isomers. Practically speaking, these solutions are most easily accessed by repeating the calculations while constraining $\beta^\text{dSCGF(n)}_2$ over a large interval, thus producing a so-called quadrupole-deformed total energy curve (TEC) whose minima deliver the physical solutions of interest~\cite{PRC}. 

\paragraph*{Results.}

\begin{figure}
	\includegraphics[scale=0.80]{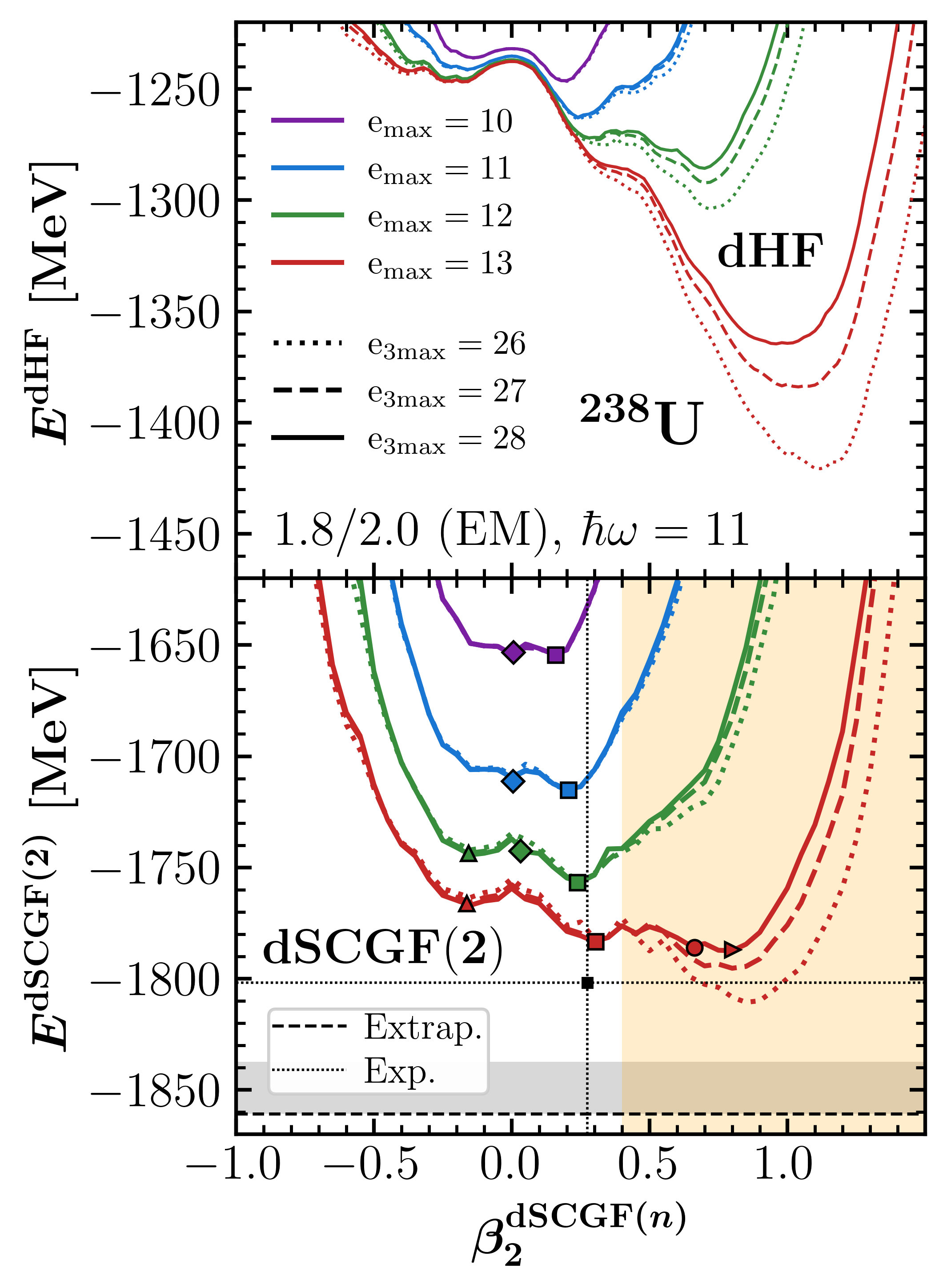}
	\caption{Deformed TEC in $^{238}$U computed with the \magic{} Hamiltonian~\cite{Hebeler11a} for $\emax=10-13$ and $\etmax=26-28$. \textit{Upper panel}: dHF results. \textit{Bottom panel}: dSCGF$(2)$ results. Full symbols report solutions obtained through unconstrained calculations for $\etmax=28$. The horizontal dotted line indicates the experimental binding energy whereas the black square on top of it further stipulates the effective quadrupole deformation extracted from data~\cite{Pritychenko:2013gwa}. The dashed line indicates the dSCGF$(3)$ energy of the first prolate minimum extrapolated to the infinite basis-size limit ($\emax \rightarrow \infty$) at fixed $\etmax=28$.  The band corresponds to the part associated with the third-order correlation energy.
    The vertical orange shaded  band marks the range of intrinsic $\beta_2$ deformation for which results are deemed unreliable for presently accessible $\emax$ and $\etmax$ values.}
	\label{Fig1main}
\end{figure}

Figure~\ref{Fig1main} displays the dHF=dSCGF($1$) and dSCGF($2$) TECs in $^{238}$U for $\emax$ varying from 10 to 13, and $\etmax$ from 26 to 28. The optimal~\footnote{The chosen sHO frequency optimizes the convergence of  dSCGF($2$) results as a function of $\emax$. The optimal frequency at the dHF level is smaller  ($\hbar \omega = 9$\,MeV). As discussed in the Supplemental Material, using the latter frequency actually amplifies the pathological behavior of the dHF TEC discussed below.} sHO frequency at the dSCGF($2$) level is employed ($\hbar \omega = 11$\,MeV). The upper panel illustrates the dubious evolution of the mean-field dHF TEC at large deformations as $\emax$ increases. While the minimum at $\beta_2\approx 0.3$  empirically expected~\cite{AMEDEEDATABASE} to describe the nuclear ground state disappears beyond $\emax = 12$, a minimum at ever increasing deformation and total energy develops beyond $\emax = 13$, without showing any sign of convergence. As $\emax$ increases, one also observes that the maximum $\etmax=28$ value that can be numerically handled at present becomes more and more insufficient to converge the dHF TEC beyond $\beta_2 \approx 0.4$. Since the 3N interaction stiffens the TEC at large deformations as $\etmax \rightarrow 3\,\emax$, the pathological behavior of the computed TEC is currently exaggerated as $\emax$ increases. The dHF TEC along with the role played by the 3N interaction and its rank reduction are analyzed in detail in the End Matter and the Supplemental Material.

\begin{figure}
	\includegraphics[scale=0.70]{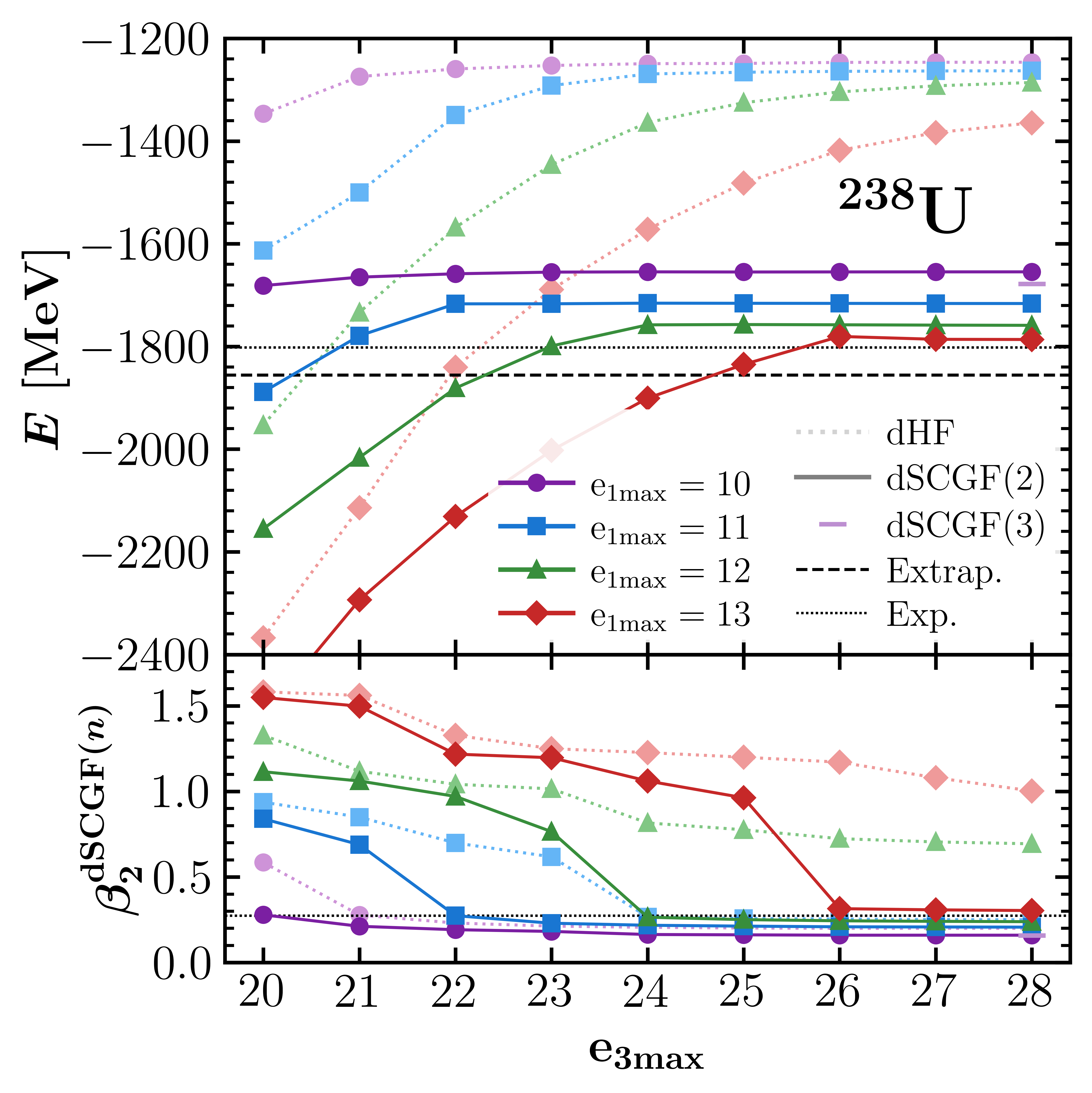}
	\caption{Energy (upper panel) and intrinsic deformation (bottom panel) of the dHF absolute minimum (colored dashed lines) and normal prolate dSCGF($2$) minimum (colored full lines) in $^{238}$U as a function of $\etmax$, for $\emax$ ranging from 10 to 13. The \magic{} Hamiltonian~\cite{Hebeler11a} is employed. The dSCGF($3$) solution at $\emax=10$ and $\etmax=28$ is also reported. 
    In the upper panel, the black-dashed line indicates the dSCGF($2$) energy of the normal prolate minimum extrapolated to the infinite basis-size limit ($\emax\rightarrow\infty$), to which is added the third-order correction  at $\emax=10$ and $\etmax=28$. The black-dotted line reports the experimental binding energy of $^{238}$U in the upper panel and the intrinsic deformation extracted from the experimental $B(E2; 0^+_1 \rightarrow 2^+_1)$~\cite{Pritychenko:2013gwa} in the lower panel.}
	\label{Fig2main}
\end{figure}

While the dubious behavior of the mean-field TEC was believed to compromise any possibility to perform meaningful {\it ab initio} calculations of heavy, intrinsically deformed, nuclei, the lower panel of Fig.~\ref{Fig1main}, along with Fig.~\ref{Fig2main}, demonstrates that it is in fact possible. Indeed, the addition of correlations beyond the deformed mean field strongly impacts the TECs. At fixed $\emax$ and $\etmax$ values, the dSCGF($2$) TEC is much stiffer than the dHF one and displays a very different topology, especially at large prolate deformations. Although on a milder scale, this key effect of correlations was already observed in $^{28}$Si~\cite{PRC}. More specifically, it is presently observed that
\begin{enumerate}
\item The number and characteristics of the minima in the dSCGF$(n)$ TEC evolve with the many-body truncation order $n$ and the $\emax$ basis truncation. While no minimum appears at normal prolate deformation ($\beta_2 \approx 0.3$) in the dHF TEC beyond $\emax=12$, such a minimum does arise in the dSCGF$(2)$ TEC. Crucially, and as best seen from Fig.~\ref{Fig2main}, the corresponding solution displays a converging pattern with $\emax$~\footnote{A converging excited dSCGF$(2)$ solution is also found on the oblate side for $\emax \geq 12$. Such a state is consistent with what is found from empirical energy density functional (EDF) calculations~\cite{AMEDEEDATABASE}. In this case, the solution does possess a corresponding minimum in the dHF TEC that is actually present already at smaller $\emax$ values.}. Up to the maximum $\emax=13$ value under consideration, the state is already well converged for $\etmax=26$. This solution is a convincing {\it ab initio} candidate for the $^{238}$U (intrinsic) ground-state based on the \magic Hamiltonian. The corresponding intrinsic deformation reported in the lower panel of Fig.~\ref{Fig2main} is compatible with the one extracted from the experimental electromagnetic transition probability $B(E2; 0^+_1 \rightarrow 2^+_1)$~\cite{Pritychenko:2013gwa}.
Its energy extrapolated to the infinite basis-size limit ($\emax \rightarrow \infty$) reported in the upper panel of Fig.~\ref{Fig2main} ($-1860.768$\,MeV) overshoots the experimental value ($-1801.696$\,MeV) by $3\%$. This result is compatible with the trend with $A$ already identified for this Hamiltonian through mid-mass nuclei~\cite{Stroberg21,Vernik2026} and with another chiral Hamiltonian up to $^{208}$Pb~\cite{Arthuis26,Hu25}. The third-order ADC($3$) contribution to the total binding energy~\footnote{The third-order correlation energy is computed at $\emax=10$ as the total energy difference $E^{\text{dSCGF}_0(3)}-E^{\text{dSCGF}_0(2)}$ and is added on top of the extrapolated dSCGF$_0(2)$ energy. This procedure delivers a converged value for this third-order correlation energy contribution~\cite{Soma20a}.} is $-23.4$\,MeV, i.e.\ $5.5\%$ of the second-order correlation energy, which is also not inconsistent with expectations~\cite{Soma20a}.   
\item In contrast, and as also shown in Fig.~\ref{Fig2main}, the dubious dHF minimum at very large deformation shows a diverging pattern with $\emax$. At the dSCGF($2$) level, two close-lying superdeformed minima appear at $\emax=13$. Their energy, deformation and $\etmax$ dependence are however much reduced compared to the dubious dHF minimum. Still, at $\emax=13$ this $\etmax$ dependence remains significant up to $\etmax=28$, i.e. comparing the TECs obtained at $\etmax=26, 27$ and $28$, one indeed observes that both dSCGF($2$) superdeformed minima are pushed up with $\etmax$. They would very probably end up above the normal-deformed minimum in the full $\etmax=39$ limit and at a smaller intrinsic deformation than at $\etmax=28$~\footnote{It is also very likely that both superdeformed solutions would merge into a single one for larger $\etmax$ values.}. At this point in time, pushing the description to even larger $\emax$ at $\etmax=28$ is however unsafe as far as the $\etmax$ convergence is concerned. It can be conjectured that, in a fully converged setting, the very deformed dSCGF$(2)$ solution would be a credible candidate for the so-called fission isomer in $^{238}$U. With current numerical capabilities, however, the part of the dSCGF TEC  covered by the orange shaded area in Fig.~\ref{Fig1main} corresponding to $\beta^{\text{dSCGF}}_2 \geq 0.4$ must be discarded as it relates to largely unconverged calculations.
\end{enumerate}

\begin{figure}
	\includegraphics[scale=0.70]{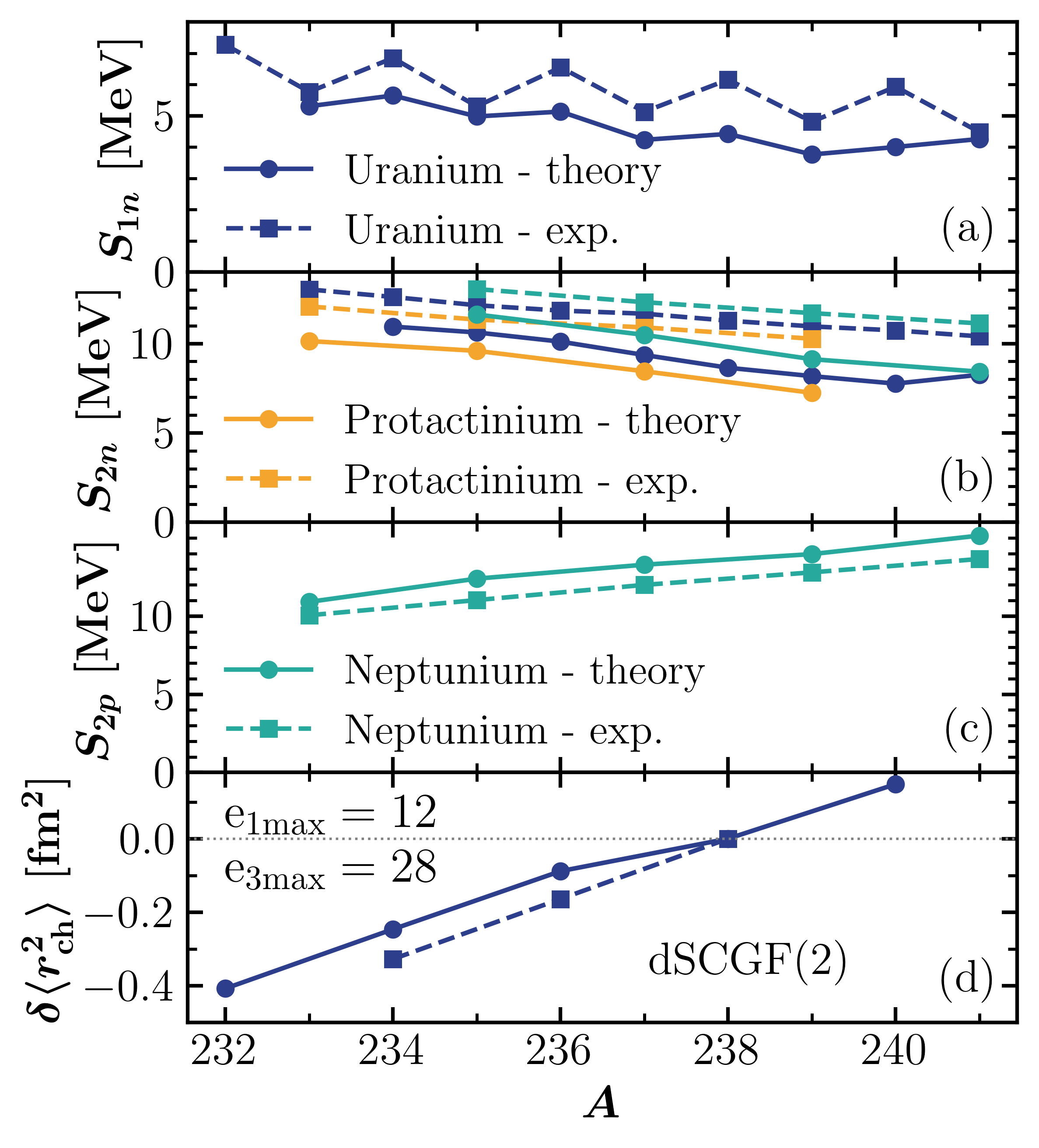}
	\caption{Theoretical (dSCGF($2$)) and experimental results in Uranium (U), Protactinium (Pa) and Neptunium (Np) isotopes. Panel (a): one-neutron separation energies in U isotopes. Panel (b): two-neutron separation energies along the three isotopic chains. Panel (c): two-proton  separation energies in Np isotopes. Panel (d) mean-square charge radius isotopic shift in U isotopes. Energies (mean-square radii) are computed using the \magic{}~\cite{Hebeler11a} ($\go{}$~\cite{Jiang20}) Hamiltonian in $\emax=12$, $\etmax=28$.}
	\label{Fig3main}
\end{figure}

With the converged {\it ab initio} $^{238}$U ground-state at hand, additional observables can be investigated~\footnote{The way separation energies, the mean-square charge radius and the charge density distribution are computed is detailed in the Supplemental Material.}. Figure~\ref{Fig3main} reports on one- and two-nucleon separation energies computed along Uranium ($Z=92$), Protactinium ($Z=91$) and Neptunium ($Z=93$) isotopic chains with the \magic{} Hamiltonian. Panel (a) shows that experimental one-neutron separation energies are reproduced in U isotopes with similar accuracies as for light and mid-mass nuclei~\cite{Stroberg21}. This is remarkable given the extraordinary extrapolation embodied by the present calculations. The too low amplitude of the odd-even staggering is also consistent with the insufficient pairing correlations displayed by {\it ab initio} calculations of mid-mass nuclei based on currently available nuclear Hamiltonians~\cite{Scalesi:2024nao,Scalesi:2026cpe}. In panel (b), while experimental two-neutron separation energies are typically underestimated by 2-3\,MeV, the trend with neutron number and the hierarchy between Pa, U and Np isotopic chains are well reproduced. In panel (c) the absolute value and trend of experimental two-proton separations along Np isotopes are remarkably well accounted for. 

As reported in the End Matter, the absolute ground-state charge radius computed with the $\go{}$ Hamiltonian underestimates the measured one~\cite{Angeli13} by only  $0.03$\,fm ($0.4\%$), which is consistent with what has been obtained up to the tin region~\cite{Demol26}.  In contrast, the charge radius delivered by the \magic{} Hamiltonian underestimates the experimental value by $0.31$\,fm ($5.4\%$), which is also in line with lighter nuclei~\cite{Lapoux16,Arthuis26,Demol26}.  Panel (d) of Fig.~\ref{Fig3main} further shows that the evolution of the mean-square charge radius computed in U isotopes with the $\go{}$ Hamiltonian is relatively well reproduced in spite of a slight kink downwards in $^{236}$U that is not visible in the experimental data. 

Last but not least, the charge density distribution is compared to two different two-point Fermi distributions fitted to elastic electron scattering data~\cite{deVries87} in the End Matter. Consistent with the charge radius, the charge density distribution obtained from the \magic{} Hamiltonian is at odds with the one extracted from experimental data, whereas the latter is well reproduced when employing the $\go{}$ Hamiltonian.

\paragraph*{Conclusions.} The description of heavy and superheavy nuclei from first principles constitutes a holy grail of nuclear theory. Building upon the tremendous extension of {\it ab initio} nuclear many-body calculations over the last fifteen years, the present work makes a huge step forward by demonstrating the possibility to perform controlled calculations of very heavy nuclei.  

Thanks to a highly efficient numerical implementation of the novel deformed self-consistent Green's function formalism, the ground-state binding energy, charge radius and charge density distribution of the iconic $^{238}$U are accessed in a controlled fashion and shown to be in fair agreement with experimental data. This is remarkable given the huge extrapolation in mass embodied by the present calculations compared to previously available ones. The study, extended to neighboring Protactinium and Neptunium isotopic chains, further shows that one- and two-nucleon separation energies computed from state-of-the-art chiral effective field theory interactions are in similar agreement with data as in much lighter nuclei. 

While the present work brings the upper-end of the nuclear chart within reach of theoretical predictions based on first-principles, it also demonstrates that accessing states in this region characterized by very large intrinsic deformations, such as the fission isomer in $^{238}$U, will require overcoming additional numerical bottlenecks.

\paragraph*{Acknowledgments.}
The authors acknowledge B.~Bally, A.~Ekstr\"om, C.~Forss\'en, and M.~Frosini for useful discussions. A.S.\ acknowledges the Swedish Research Council (Grants No.~2021-04507 and No.~2025-05618) and the National Academic Infrastructure for Supercomputing in Sweden (NAISS), funded by the Swedish Research Council, for providing computational resources.
This project was also provided with HPC and storage resources by GENCI at TGCC, France, thanks to Grant No. A0190513012 on the supercomputer Joliot-Curie’s ROME partition.

\bibliography{biblio}

\clearpage

\section*{End Matter}

This section presents a more detailed analysis of the mean-field TEC and extends the $^{238}$U study to the mean-square charge radius and the charge density distribution.
For the latter two the $\go{}$ Hamiltonian~\cite{Jiang20} is also used given that the \magic{} Hamiltonian, while delivering accurate energies, is known to strongly underestimate charge radii in light and medium-mass nuclei~\cite{Lapoux16,Arthuis26,Demol26}.

\begin{figure*}
	\includegraphics[scale=0.9]{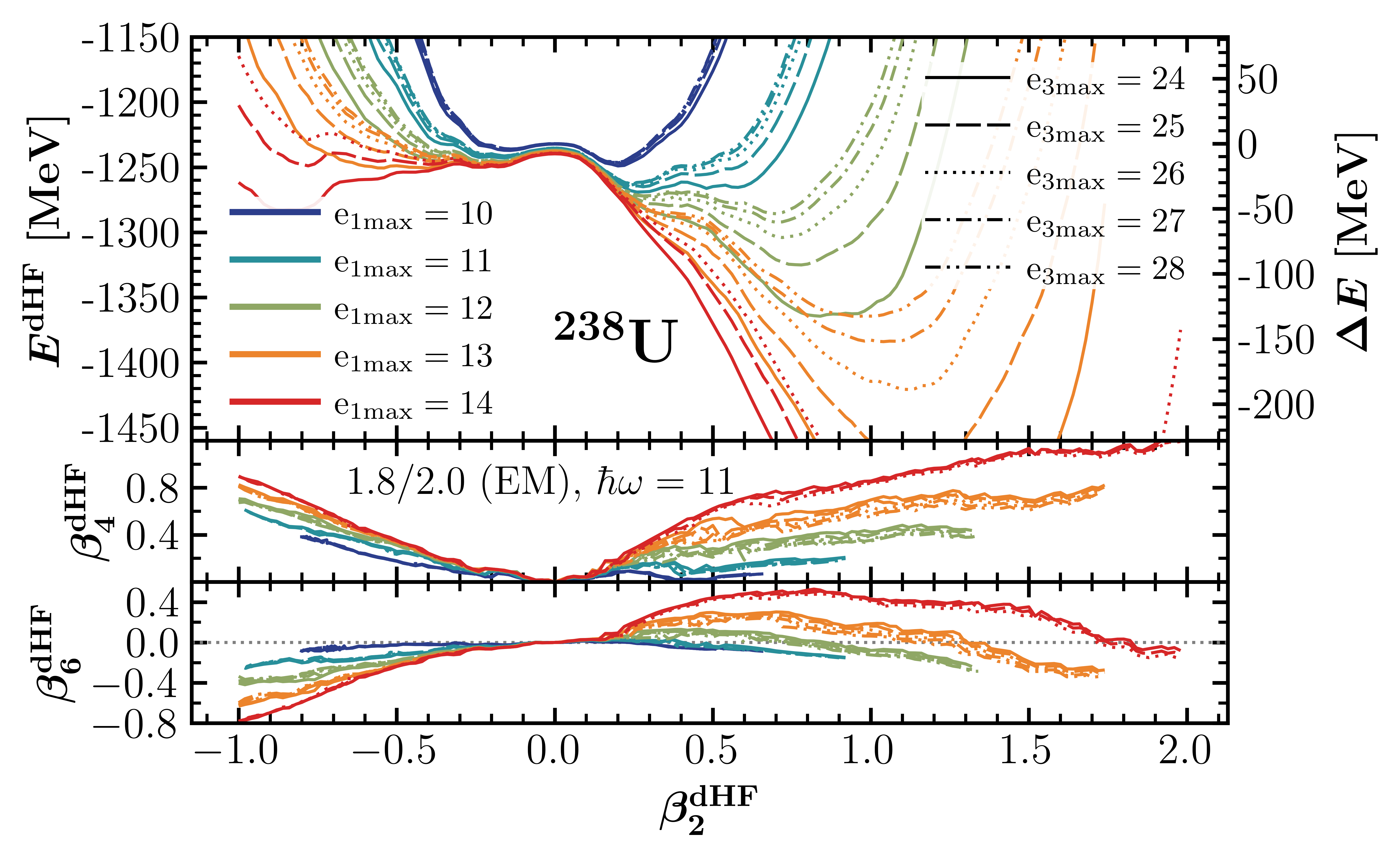}
	\caption{\emph{Upper panel}: dHF TEC of $^{238}$U for several values of $\emax$ and $\etmax$.
    The right axis shows the energy relative to the $\beta_2^{\rm dHF}=0$ point of the $\emax=10$, $\etmax=28$ curve, so that the overall lowering of the other TECs can be read off directly.
    \emph{Middle} and \emph{lower panels}: $\beta_4^{\rm dHF}$ and $\beta_6^{\rm dHF}$ deformations as a function of $\beta_2^{\rm dHF}$, respectively.}
	\label{fig:TEC_multipoles}
\end{figure*}

\paragraph*{Behavior of the mean-field TEC.}
To further investigate the unconventional behavior of the dHF TEC at large $\beta_2^{\rm dHF}$ discussed in the main text, two additional deformation parameters are introduced: the hexadecapole $\beta_4^{\rm dHF}$ and the hexacontatetrapole $\beta_6^{\rm dHF}$.
They are obtained in the same way as $\beta_2^{\rm dHF}$, by contracting the multipole moment operator $Q_{\lambda 0}$ ($\lambda=4,6$) with the dHF one-body density matrix, as detailed in Ref.~\cite{PRC}.
The upper panel of Fig.~\ref{fig:TEC_multipoles} displays the dHF TEC for several  $\emax$ and $\etmax$ values whereas the middle and lower panels provide the corresponding $\beta_4^{\rm dHF}$  and $\beta_6^{\rm dHF}$ values, respectively.

At $\emax=10$ the dHF TEC is essentially converged with respect to $\etmax$ and is qualitatively similar to the one obtained from phenomenological energy density functional (EDF) calculations~\cite{AMEDEEDATABASE} even though the absolute minimum arises at smaller deformation ($\beta_2^{\rm dHF}\approx 0.2$).
Moving to $\emax=11$, the TEC displays a second prolate minimum at large deformation ($\beta_2^{\rm dHF}\approx 0.5$) for $\etmax=24$ that is lying slightly above the first minimum. Raising $\etmax$ to 28 stiffens the converged TEC in which the second minimum has disappeared. At the same time the first minimum has moved to slightly larger deformation than for $\emax=10$.
At $\emax=12$ and $\etmax=24$, the TEC presents a single deep minimum at very large deformation. Raising $\etmax$ stiffens the dHF TEC that is essentially converged at  $\etmax=28$. In the latter the first prolate minimum is now very shallow whereas the one at large deformation has become the absolute minimum. At $\emax=13$ and $14$, no minimum at all appears at moderate prolate deformation. At the same time, the global minimum continues to drop to even more unreasonable energies and to shift to larger and larger $\beta_2^{\rm dHF}$. Correspondingly, $\etmax$ values required to deliver a converged dHF TEC enter a more and more inaccessible range such that the dubious topology appearing in Fig.~\ref{fig:TEC_multipoles} is artificially exaggerated.

As can be empirically appreciated from the middle and lower panels of Fig.~\ref{fig:TEC_multipoles}, the collapse of the dHF TEC correlates to a large extent with $\beta_4^{\rm dHF}$ and $\beta_6^{\rm dHF}$ reaching unexpectedly large values.
The hexadecapole degree of freedom constitutes a next-to-leading-order effect in the multipole expansion of the nuclear shape such that $\beta_4$ is generically expected to be significantly smaller than $\beta_2$~\cite{Hendrie68,Bemis73}.
This hierarchy is confirmed experimentally, e.g., by Coulomb-excitation and muonic x-ray determinations of the $E2$ and $E4$ moments in the actinides~\cite{Bemis73,Zumbro84}. It is also confirmed theoretically via large-scale EDF surveys across the nuclear chart~\cite{NithishKumar23}, where the largest ground-state hexadecapole deformations remain well below their quadrupole counterparts. On the oblate side the collapse of the dHF TEC is less dramatic, which is consistent with $\beta_4^{\rm dHF}$ remaining smaller than at the corresponding prolate point.

This latter behavior can be traced back to the simple geometric identity relating $\beta_4$ to $\beta_2$
\begin{equation}
\beta_4(\beta_2) \approx a\,\beta_2^2 + b\,\beta_2^3 + \mathcal{O}(\beta_2^4),
\label{eqn:beta4}
\end{equation}
where the derivation of the coefficients $a>0$ and $b>0$ is detailed in the Supplemental Material. While the leading term is quadratic in $\beta_2$, the cubic term leads to $\beta_4(-|\beta_2|) < \beta_4(|\beta_2|)$, so that $\beta_4$ is smaller on the oblate side, which is consistent with the reduced collapse observed on that side of the dHF TEC.

Consistent with the above analysis, producing the dHF TEC while further constraining $\beta_4^{\rm dHF}$ and $\beta_6^{\rm dHF}$ to zero is found to effectively suppress the collapse for all $\emax$ and $\etmax$ values, thus yielding a TEC free of the dubious low-energy minimum at very large deformation.

\begin{figure}
	\includegraphics[scale=0.93]{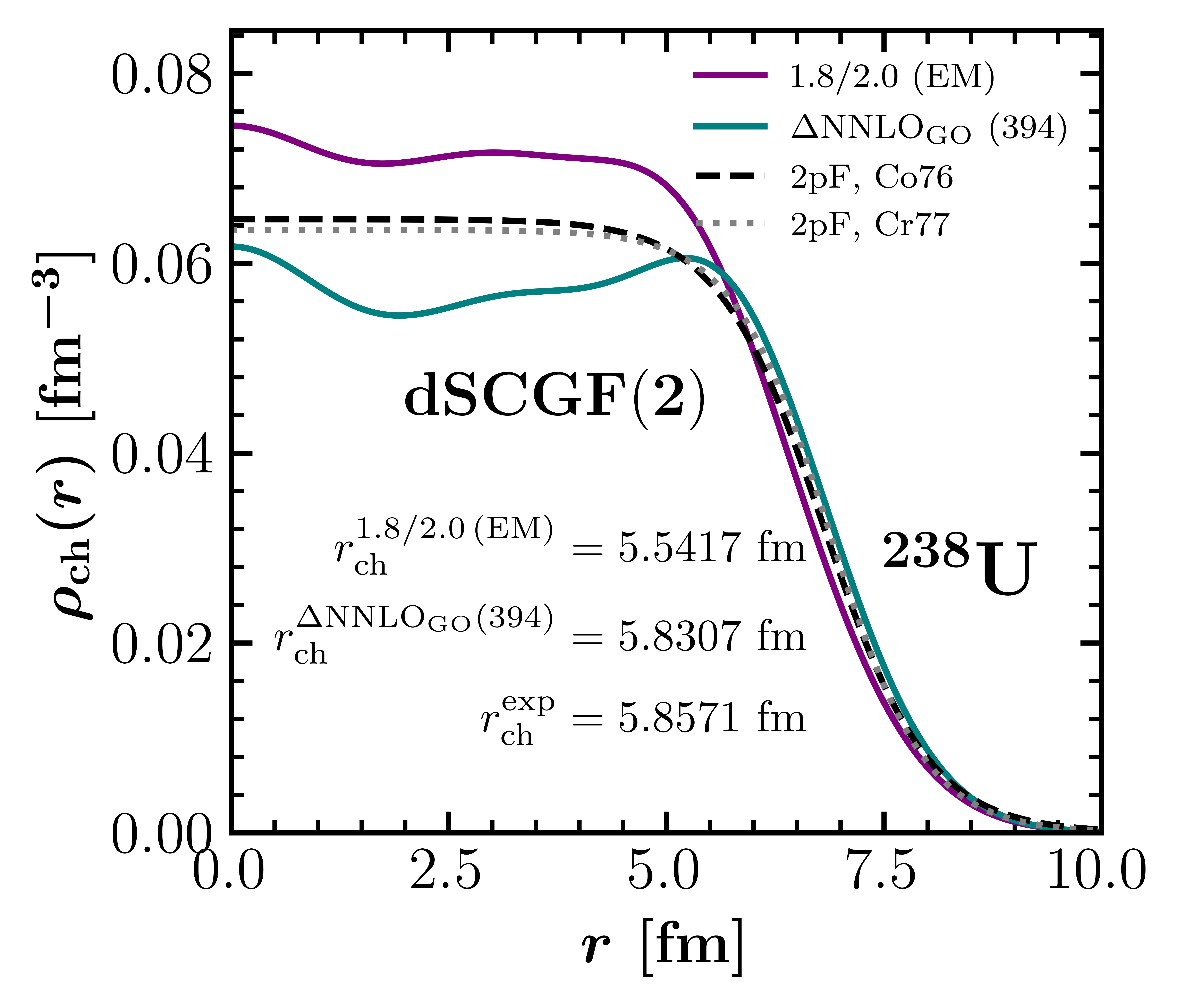}
	\caption{Ground-state charge density distribution of $^{238}$U obtained from dSCGF($2$) calculations based on the \magic{} and $\go{}$ Hamiltonians using $\emax=12$ and $\etmax=28$.
	Results are compared to two two-point Fermi distributions fitted to elastic electron scattering data~\cite{deVries87}. The figure also compares the ground-state experimental charge radius to the one obtained from dSCGF($2$) calculations using both nuclear Hamiltonians.}
	\label{fig:CDD}
\end{figure}

\paragraph*{Charge radius and density distribution.}

While the study presented in this Letter has so far been restricted to total energies obtained with the \magic{} Hamiltonian, the present section examines the theoretical results on the ground-state root mean-square charge radius and charge density distribution obtained using both the $\go{}$ and \magic{} Hamiltonians.  Details on the computation of these quantities can be found in the Supplemental Material.

Figure~\ref{fig:CDD} compares the ground-state charge density distribution obtained from dSCGF($2$) calculation to two empirical two-point Fermi (2pF) profiles~\cite{deVries87}.
The dSCGF results exhibit quantum oscillations in the region $r = 0$--$5$\,fm that are absent by construction from the empirical curves based on a 2pF parametrization that cannot accommodate such fluctuations.

The two Hamiltonians induce markedly different behaviors, the \magic{} charge density remaining well above the $\go{}$ and the empirical ones up to $r = 6$\,fm before diving under them at larger radii. This is consistent with the known tendency for the \magic{} Hamiltonian to deliver excessively small charge radii~\cite{Lapoux16,Arthuis26,Demol26}, which are nothing but the second moment of the charge density distribution. Indeed, this tendency extends all the way to $^{238}$U where the ground-state charge radius delivered by the \magic{} Hamiltonian underestimates the experimental value~\cite{Angeli13} by $0.31$\,fm ($5.4\%$). 

At $r = 0$\,fm, the charge density distribution obtained from the $\go{}$ Hamiltonian lies significantly closer to the empirical profile than the one generated by the \magic{} Hamiltonian that overshoots it. This closer agreement also extends to large radii.  Correspondingly the ground-state charge radius computed from the $\go{}$ Hamiltonian only underestimates the measured one by about $0.03$\,fm ($0.4\%$), which is also consistent with what was obtained up to the tin region~\cite{Demol26}.

\clearpage

\begin{widetext}
\begin{center}
{\large\bfseries SUPPLEMENTAL MATERIAL\\[6pt]}
\end{center}
\end{widetext}
\vspace{6pt}

This Supplemental Material collects the analyses supporting the $^{238}$U results reported in the Letter.
The impact of three-nucleon forces at the deformed Hartree-Fock level is first quantified.
The convergence of the deformed self-consistent Green's function solutions with respect to the basis-size parameters $\emax$ and $\etmax$ is then examined in detail.
A subsequent section documents how the observables presented in the main text are computed.
A final section derives the geometric origin of the $\beta_4(\beta_2)$ relation utilized in the End Matter to interpret the behavior of the deformed Hartree-Fock  total energy curve at large deformation.

\section{$^{238}$U analysis}
\label{sec:U238}

Performing a similar analysis to the one conducted in Ref.~\cite{PRC} for $^{28}$Si, the convergence of dHF and dSCGF calculations of $^{238}$U is characterized. The numerical challenge that compounds such an effort is huge~\cite{PRC}, in particular due to the large $\emax$ and $\etmax$ values necessary to produce converged results in very heavy nuclei.

\subsection{Impact of the 3N interaction on the dHF TEC}

\begin{figure}
	\includegraphics[scale=0.60]{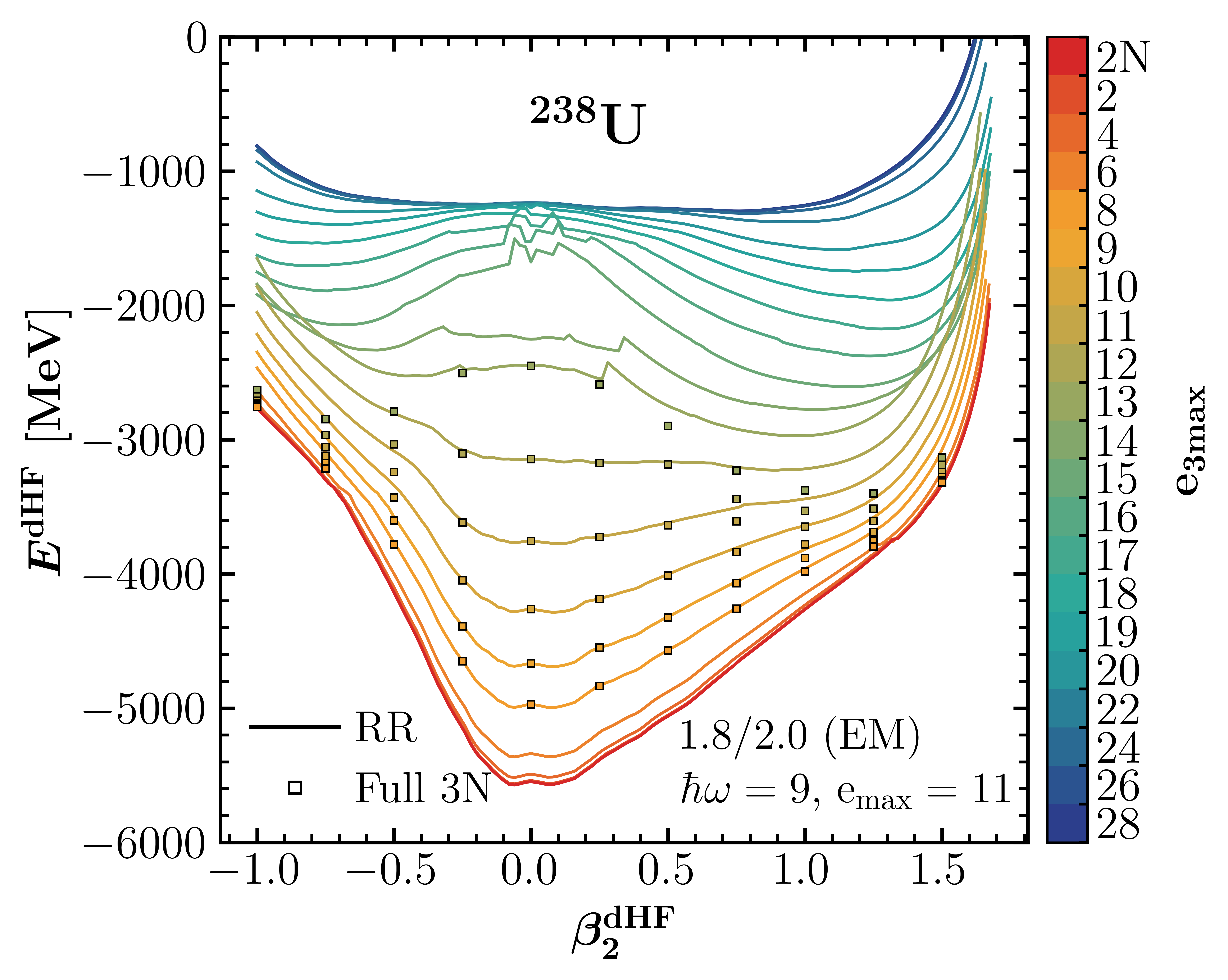}
	\caption{Deformed Hartree-Fock TEC of $^{238}$U computed at $\emax=11$ with the \magic~\cite{Hebeler11a} Hamiltonian for different $\etmax$ truncations based on the optimal sHO frequency ($\hbar \omega = 9$\,MeV) at the dHF level. Results are based on the RR form of the 3N interaction but are benchmarked against those obtained using the explicit 3N operator for $\etmax=8$ to $14$.}
	\label{TEC238U}
\end{figure}

Figure~\ref{TEC238U} displays the dHF TECs for $^{238}$U using $\emax=11$ and $\etmax$ varying between $0$ and $28$, the latter being close to but not exactly reaching $3\,\emax$. The optimal sHO frequency at the dHF level ($\hbar \omega = 9$\,MeV) is employed.

\begin{figure}
	\includegraphics[scale=0.60]{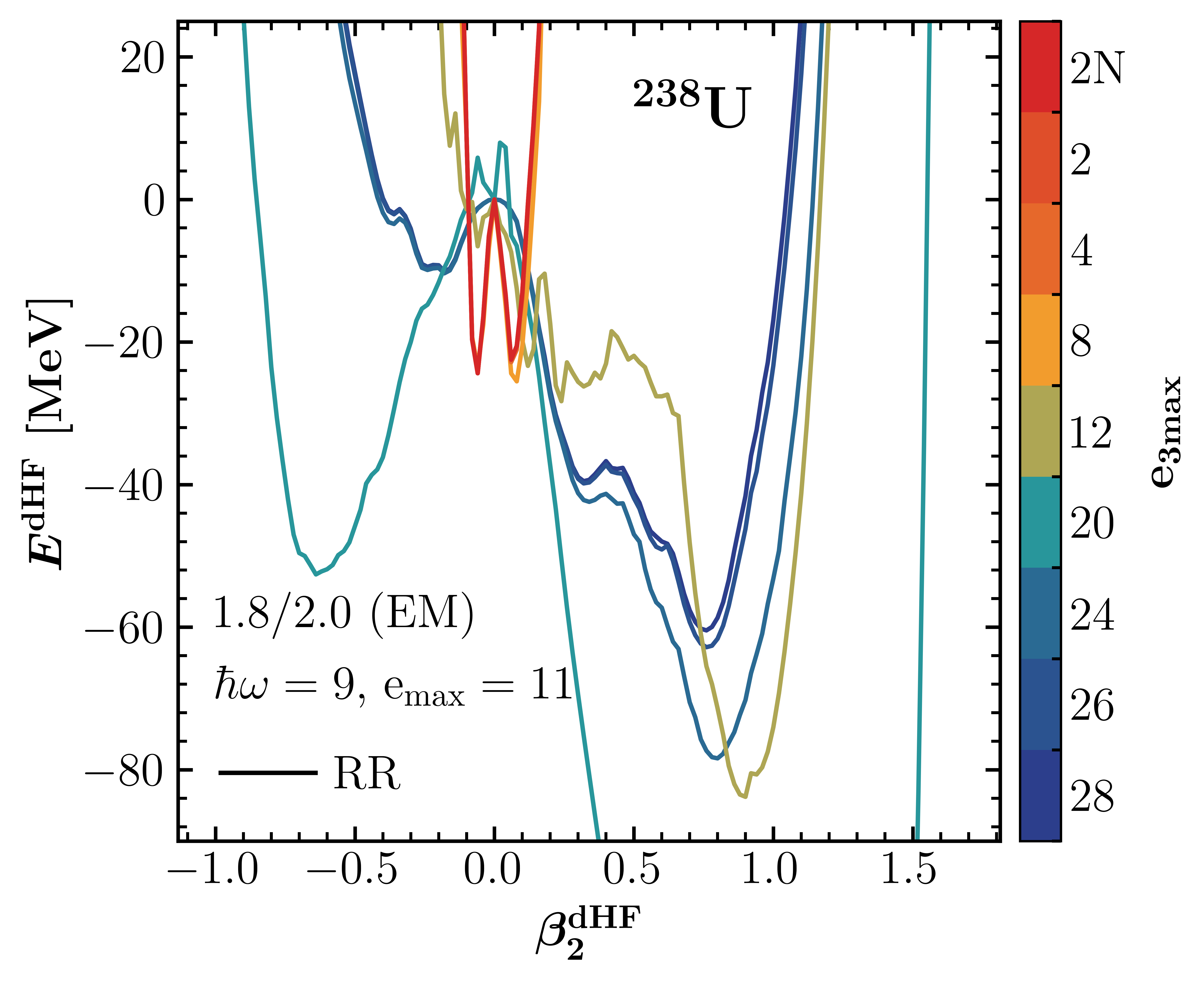}
	\caption{Same as Fig.~\ref{TEC238U} but with energies shown relative to their value at $\beta^{\rm dHF}_2=0$.}
	\label{TEC238Urescaled}
\end{figure}

Starting with the benchmark of the RR against the full 3N interaction operator, calculations with the latter can only be performed up to $\etmax=14$, which is far from $3\,\emax=33$. Still, one observes the same trend as in~\cite{PRC} for $^{28}$Si: the RR approximation is excellent except at large prolate deformation $\beta^{\text{dHF}}_2$ for the intermediate $\etmax$ values that can be presently employed. More specifically, the full 3N operator and its RR approximation both deliver TECs displaying a very deep minimum at large prolate deformation, the former generating an even deeper minimum than the latter. As $\etmax$ further increases towards $3\,\emax$, the TEC delivered by the RR stiffens again at large deformations such that a troublesome deep minimum eventually remains but on a much smaller scale as can be better appreciated from Fig.~\ref{TEC238Urescaled}. Based on the analysis performed in  $^{28}$Si~\cite{PRC}, one can speculate that the TEC produced by the full 3N interaction operator eventually `catches up' with the one obtained based on the RR approximation such that both essentially agree in the complete $\etmax=3\,\emax$ limit.

The topology of the TEC obtained for $\etmax=28$ results from the effect of adding progressively the 3N interaction on top of the 2N interaction. The latter alone delivers a TEC that is completely at odds with the one typically obtained from EDF calculations and with the expected phenomenology of $^{238}$U. As better seen in Fig.~\ref{TEC238Urescaled} the corresponding TEC displays two pronounced oblate and prolate minima at small intrinsic deformations  $\beta^{\text{dHF}}_2 \approx \pm 0.1$ followed by an extremely stiff TEC at larger deformations. Progressively including the effect of the 3N interaction as a function of $\etmax$ provides a strongly repulsive effect that first affects small deformations, creating an extraordinarily deep minimum  at large prolate deformation beyond $\etmax=12$. As discussed above and as already seen in $^{28}$Si~\cite{PRC}, although on a much smaller scale, further increasing $\etmax$ eventually impacts larger deformations more strongly in such a way that the catastrophic character of the minimum at large prolate deformation is largely reduced. As visible from  Fig.~\ref{TEC238Urescaled}, one is left with an essentially converged TEC with respect to $\etmax$ that varies on a much more reasonable scale than without 3N interaction but that is still at odds with the TEC obtained from EDF calculations. Indeed, the latter typically displays a prolate minimum at $\beta_2\approx 0.3$ identified with the (intrinsic) $^{238}$U ground state and an excited local minimum at $\beta_2\approx 0.5$ identified with the (intrinsic) fission isomer. While a local minimum is observed at $\beta^{\text{dHF}}_2 \approx 0.3$ in Fig.~\ref{TEC238Urescaled} for the converged TEC, it is largely superseded by a much deeper minimum at  $\beta^{\text{dHF}}_2 \approx 0.8$. Increasing $\emax$ further while keeping $\etmax$ fixed at 28, which is the largest value that can be presently handled numerically, actually leads to a disappearance of the local minimum at $\beta^{\text{dHF}}_2 \approx 0.3$ and to a continuous deepening of the absolute minimum that is continuously pushed to larger prolate deformation. The latter is of course due, for an increasing part, to the fact that the fixed $\etmax=28$ value at hand lies further and further away from the complete $3\,\emax$ value as $\emax$ increases beyond $9$. Consequently, it is premature at this point in time to speculate on the reality of the evolution of the dHF TEC with $\emax$ at very large deformations\footnote{Future studies of this highly deformed portion of heavy nuclei's TECs will also have to assess the impact of the truncation of total angular momentum components in 2N and 3N interaction matrix elements prior to their transformation in the sHO basis.}.

The unexpected topology of the dHF TEC identified above is not unique to $^{238}$U but rather constitute an extreme version of the patterns developing continuously with the increasing nuclear mass. In this respect, the present study constitutes a worst-case scenario given that, while the phenomenologically expected intrinsic deformation of the $^{238}$U ground state is large ($\beta_2\approx 0.3$), the unorthodox behavior of the dHF TEC extends to smaller deformations than in lighter heavy nuclei in such a way that even the minimum associated with the ground state is impacted and actually disappears at larger $\emax$ as discussed in the following section. 

\subsection{dSCGF$_0$($2$) solutions}

\begin{figure}
	\includegraphics[scale=0.75]{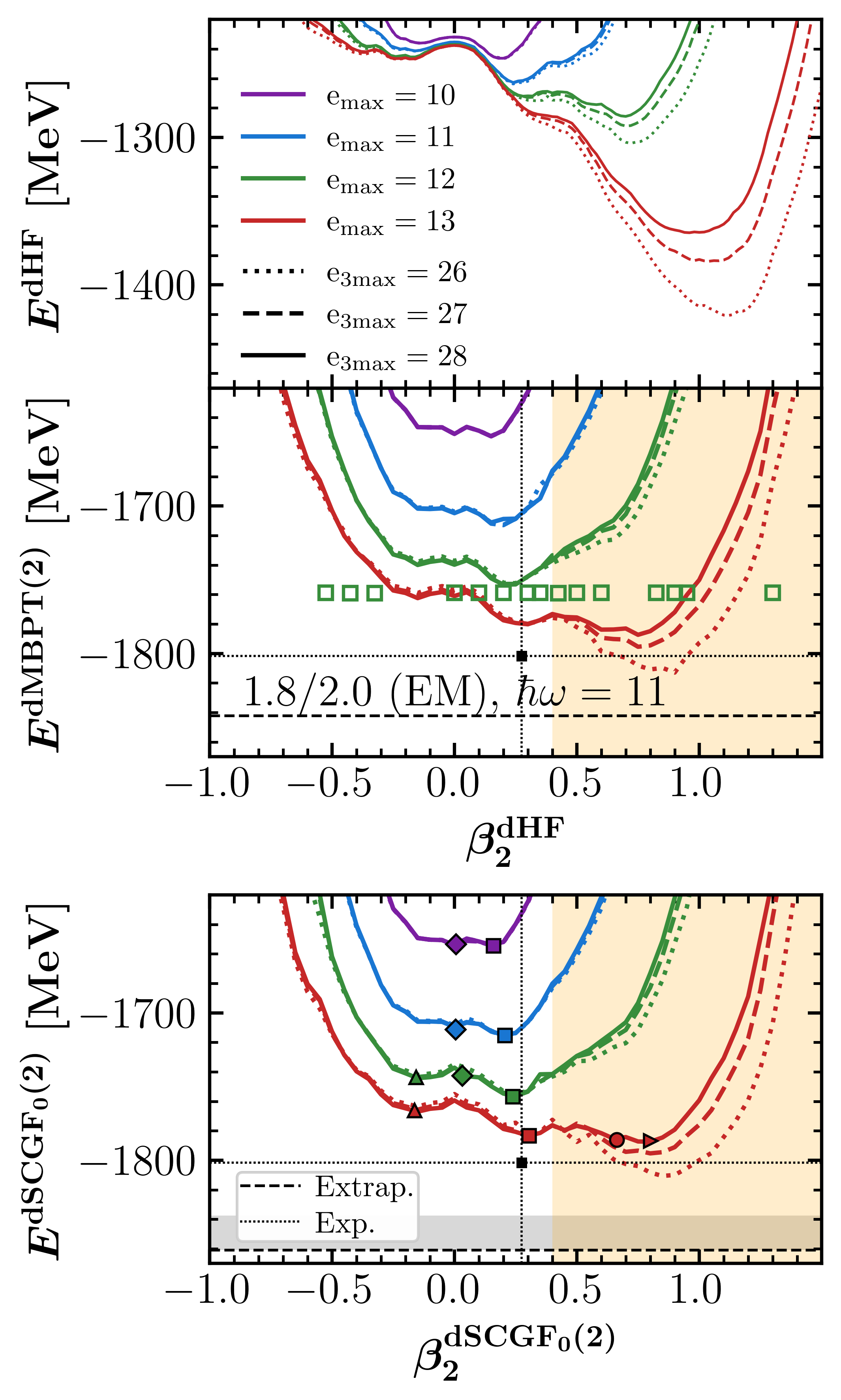}
	\caption{Axial quadrupole TECs in $^{238}$U computed for $\emax=10-13$ and $\etmax=26-28$ using $\hbar\omega=11$\,MeV and the \magic Hamiltonian. \textit{Upper panel}: dHF TECs against $\beta_2^{\text{dHF}}$. \textit{Middle panel}: TECs from unconstrained dMBPT(2) calculations performed on top of the dHF TECs. Empty symbols report results from unconstrained  dSCGF$(2)$ calculations performed on top of the dHF TEC for $\emax=12$ and $\etmax=28$. \textit{Bottom panel}: dSCGF$_0(2)$ TECs against $\beta^{\text{dSCGF}_0(2)}_2$. Full symbols report solutions obtained through unconstrained calculations for $\etmax=28$.  In the middle and bottom panels, the horizontal dotted line indicates the experimental binding energy whereas the black square on top of it further stipulates the effective quadrupole deformation extracted from data~\cite{Pritychenko:2013gwa}. The dashed line in the middle panel indicates the dMBPT$(2)$ energy of the first prolate minimum extrapolated to the infinite basis-size limit ($\emax \rightarrow \infty$) at fixed $\etmax=28$. The dashed line in the bottom panel does the same for the dSCGF$_0(3)$ energy where the band corresponds to the third-order correlation energy contribution obtained as the total energy difference $E^{\text{dSCGF}_0(3)}-E^{\text{dSCGF}_0(2)}$ computed at $\emax=10$ and added on top of the extrapolated dSCGF$_0(2)$ energy.
    In the middle and bottom panels, the orange shaded  band marks the range of intrinsic deformations for which results are deemed unreliable given that convergence cannot be established for presently accessible $\emax$ and $\etmax$ values.}
	\label{SCGFU238}
\end{figure}

In the following, the notation dSCGF$_0(2)$ denotes dSCGF(2) calculations for which self-consistently is only reached with respect to first-order diagrams, i.e.\ second-order contributions to the ADC($2$) matrix are computed only once from the dHF propagator and frozen throughout the self-consistent iterations~\cite{PRC}. While reducing very significantly the computational cost of the extensive calculations presented below, this partial self consistency does not impact the qualitative behavior of the results and affect them quantitatively only moderately\footnote{While this is the case when low-momentum Hamiltonians are employed, as the \magic of the present study, it might not apply to harder interactions, for which fully self-consistent calculations are typically necessary.}.

With the above analysis of the dHF TEC at hand, dSCGF$_0(2)$ calculations are performed at the optimal $\hbar \omega = 11$\,MeV for $\emax$ varying from 10 to 13, and $\etmax$ from 26 to 28. The results of three different sets of calculations are actually compared in Fig.~\ref{SCGFU238}. The upper panel displays the dHF TECs computed for $\hbar \omega = 11$\,MeV. The middle panel displays deformed many-body perturbation theory (dMBPT) TECs obtained via {\it unconstrained} calculations, i.e.\  the results of straight dMBPT calculations performed on top of each state along the dHF TECs are plotted against the starting  $\beta^{\text{dHF}}_2$ deformation. For $\emax=12$ and $\etmax=28$, this panel also shows results from unconstrained dSCGF$(2)$ calculations (empty symbols). Finally, the bottom panel of Fig.~\ref{SCGFU238} reports the dSCGF$_0(2)$ TECs against $\beta^{\text{dSCGF}_0(2)}_2$ along with the complete set of unconstrained dSCGF$_0(2)$ solutions (full symbols) that appear to be nothing but the minima of the constrained dSCGF$_0(2)$ TECs~\cite{PRC}. 

Starting with the upper panel, one observes that employing the optimal sHO frequency at the dSCGF($2$) level mitigates the unexpected patterns identified in Fig.~\ref{TEC238Urescaled} for the dHF TEC computed at $\hbar \omega = 9$\,MeV but does not remove them, e.g.\ while the disappearance of the local minimum at $\beta^{\text{dHF}}_2 \approx 0.3$ is shifted from $\emax=12$ to $\emax=13$, the appearance of a deeper minimum at very large deformation is shifted to $\emax=12$.  

Using dMBPT($2$) as the simplest method to add dynamical correlations, the middle panel illustrates how doing so strongly impacts the topology of the TECs. At fixed $\emax$ and $\etmax$ values, the dMBPT($2$) TEC is much stiffer than the dHF one and displays a very different topology, especially at large prolate deformations. Even though on a much smaller scale, this behavior was already observed in $^{28}$Si~\cite{PRC}. This key role of correlations beyond the deformed mean field is confirmed in the bottom panel where dSCGF$_0(2)$ TECs are very similar to the dMBPT(2) ones in spite of being typically shifted down by a few MeVs. 

Let us make a digression to better appreciate how unconstrained and constrained calculations work for self consistent versus non self-consistent methods. To do so, using $\emax=12$ and $\etmax=28$ as an arbitrary example, the middle panel of Fig.~\ref{SCGFU238} displays results from both unconstrained dSCGF$(2)$ (empty symbols) and dMBPT($2$) calculations. On the one hand, unconstrained dMBPT($2$) solutions keep a strong memory of the dHF starting points, thus generating a genuine TEC as a function of $\beta^{\text{dHF}}_2\approx \beta^{\text{dMBPT}(2)}_2$. On the other hand, unconstrained dSCGF($2$) calculations lose the memory of the starting dHF state through the self-consistent iterations to converge to the same solution over a large range of initial intrinsic deformations. As visible from the bottom panel, the few solutions emerging from such unconstrained calculations are nothing but the minima of the constrained dSCGF$_0(2)$ TECs\footnote{As a matter of fact, the empty symbols in the middle panel are from fully self-consistent dSCGF($2$) calculations and are thus located about 3\,MeV below the full symbols appearing as minima of the dSCGF$_0(2)$ TEC in the lower panel.}. While a self-consistent method such as SCGF has the capacity to access the physical solutions, i.e. the minima of the TEC, essentially independently of the starting point, it however requires to perform {\it constrained} calculations to extract the stiffness of the energy against intrinsic deformation. 

Let us now go back to the impact of dynamical correlations beyond the deformed mean-field that leads to crucial consequences, i.e.,
\begin{enumerate}
\item The topologies of the dHF and dSCGF$_0(2)$ TEC are largely different. 
\item The number and characteristics of the minima in the dSCGF$_0(2)$ TEC evolve as a function of $\emax$ as long as calculations are not converged. As discussed below, calculations may be (essentially) converged for a certain portion of the TEC but not for another. While this aspect appears to be far more critical in the context of present \ai{} calculations, it is well known to EDF practitioners using the sHO basis.
\item While no minimum appears at normal prolate deformation ($\beta_2 \approx 0.3$) in the dHF TEC beyond $\emax=12$, such a minimum does appear in the dSCGF$_0(2)$ TEC. Crucially, the corresponding solution displays a converging pattern with $\emax$\footnote{A converging excited dSCGF$_0(2)$ solution is also found on the oblate side for $\emax \geq 12$. Such a state is consistent with what is found from EDF calculations~\cite{AMEDEEDATABASE}. In this case, the solution does possess a corresponding minimum in the dHF TEC that is actually present already at smaller $\emax$ values.}. Up to the maximum $\emax=13$ value under consideration, the state is converged with respect to $\etmax$ at the maximum available value $\etmax=28$. This solution is a convincing \ai{} candidate for the $^{238}$U (intrinsic) ground-state based on the \magic \, Hamiltonian. The corresponding intrinsic deformation is compatible with the one extracted from the experimental $B(E2; 0^+_1 \rightarrow 2^+_1)$~\cite{Pritychenko:2013gwa}.
Its extrapolated energy to the infinite basis-size limit, including third-order corrections ($-1860.768$\,MeV)\footnote{The values of the extrapolated dMBPT(2) and dSCGF$_0$(2) energies are respectively -1842.2 $\pm$ 8.3 and -1837.4 $\pm$ 2.1\,MeV, where the errors are $1\sigma$ uncertainties extracted from the covariance matrix resulting from the fit procedure. Such uncertainty does not allow to determine the relative position of the two minima. The extrapolated dSCGF(2) energy is -1844.6 $\pm$ 3.6\,MeV, a few MeV below the dSCGF$_0$(2) one.}, overshoots the experimental value ($-1801.696$\,MeV) by $3\%$. This result is compatible with the trend as a function of the nuclear mass already identified for this Hamiltonian through mid-mass nuclei~\cite{Stroberg21,Vernik2026} and with another chiral Hamiltonian up to $^{208}$Pb~\cite{Arthuis26,Hu25}. The third-order correction to this total binding energy\footnote{The third-order correlation energy is computed at $\emax=10$ as the total energy difference $E^{\text{dSCGF}_0(3)}-E^{\text{dSCGF}_0(2)}$ and is added on top of the extrapolated dSCGF$_0(2)$ energy. This procedure delivers a converged value for this third-order correlation energy contribution~\cite{Soma20a}.} is $-23.4$\,MeV, i.e.\ $5.5\%$ of the second-order correlation energy, which is also not inconsistent with expectations~\cite{Soma20a}.   
\item While a dubious minimum at {\it very} large deformation appears at $\emax=12$ in the dHF TEC, it  only does so at $\emax=13$ in the dSCGF$_0(2)$ TEC. In fact, two close-by superdeformed solutions are found. At fixed $\emax$, e.g.\ 13, the deformation and energy of the dubious minima are much larger in the dHF TEC than in the dSCGF$_0(2)$ one. Additionally, the $\etmax$ dependence is also much more pronounced at the dHF level than at the dSCGF$_0(2)$ one. Still, for this largest available $\emax=13$ value, the $\etmax$ dependence of the dubious solutions around $\beta^{\text{dSCGF}_0(2)}_2 \approx 0.7-0.9$ remains significant up to the maximum available $\etmax=28$ value. Comparing the TECs obtained at $\etmax=26, 27$ and $28$, one indeed observes that these solutions are pushed up with $\etmax$ and would, with no doubt, end up above  the normal-deformed minimum in the full $\etmax=39$ limit and at a smaller intrinsic deformation  than at $\etmax=28$. It is also possible that only one such very deformed solution would survive. Pushing the description to even larger $\emax$ is completely unsafe from the point of view of the $\etmax$ dependence at this point in time. It can be conjectured that, in a fully converged setting, the very deformed dSCGF$_0(2)$ solution would be a credible candidate for the so-called fission isomer in $^{238}$U. With current numerical capabilities, however, it is clear that the part of the dSCGF TEC located at $\beta^{\text{dSCGF}}_2 \geq 0.4$ must be discarded given that it relates to largely unconverged calculations. The corresponding unsafe portion of the TEC appears below an orange shaded area in Fig.~\ref{SCGFU238}.
\end{enumerate}

\subsection{Conclusions}

The above analysis allows us to draw several important lessons regarding \ai{} calculations of very heavy doubly open-shell nuclei doable today.
\begin{enumerate}
\item Self-consistent non-perturbative correlation-expansion methods such as dSCGF constitute powerful theories delivering (approximate) solutions that are independent of the starting dHF states at fixed $(\hbar\omega,\emax,\etmax)$ values. 
\item Performing constrained calculations at the correlated level, e.g.\ the production of axially deformed  dSCGF($n$) TECs, is very useful to identify, characterize and follow correlated solutions as a function of basis-size parameters since the latter  appear as minima in the TEC. The prior use of significantly less costly unconstrained dMBPT($2$) calculations on top of the dHF TEC can be useful to identify minima that bear the chance to deliver converged physical solutions.
\item The addition of dynamical correlations beyond the deformed mean-field is absolutely key to reach physical solutions whose characteristics may not be easily anticipated from the deformed mean-field solutions (i.e.\ from minima in the dHF TEC).
\item Still, the convergence of dSCGF($n$) calculations with respect to  basis-size parameters (i.e.\ $\emax$ and $\etmax$ values at the optimal $\hbar\omega$ value) depends on the particular solution under consideration. In particular, solutions carrying a very large intrinsic deformation converge significantly more slowly than those carrying smaller, e.g.\ standard, intrinsic deformations. While the latter can be essentially converged\footnote{This statement means that the solution displays a converging pattern as a function of $\emax$ and $\etmax$ for values that can be handled such that a controlled extrapolation to the infinite basis-size limit can be performed. In a more refined setting, one can actually rather extrapolate to the basis dimension for which the basis-size error is of the order of the many-body truncation uncertainty~\cite{Zurek:2026xsu}.}  in a nucleus as heavy as $^{238}$U with present capabilities, the former is currently out of reach even with the most advanced \ai{} calculations presented in this work.
\item A normal-deformed dSCGF($n$) solution can exist and be essentially converged even if (i) the converged dHF TEC does not display any minimum over the range of deformations where that dSCGF($n$) solution is located and (ii) the dHF TEC is still largely unconverged at larger deformations. Consequently, one cannot judge the existence and convergence of correlated solutions based on the topology displayed by either the converged or unconverged parts of the dHF TEC. In particular, one must not disqualify the accessibility of a certain number of controlled solutions based on the fact that part of the dHF TEC at large deformations is unconverged and display an apparent pathological behavior.
\item The current situation regarding the impossibility to safely extract states characterized by very large intrinsic deformations, such as the fission isomer in $^{238}$U, calls for calculations at larger $\etmax$ values or rather calls for more efficient ways to approximate the set of three-body matrix elements retained in the first place to perform the rank reduction of the three-body operator. Pushing calculations to larger $\emax$ values must also be pursued. Eventually, \ai{} calculations may have to be performed based on a two-center harmonic oscillator basis in order to address physics associated with large intrinsic deformations~\cite{S_nchez_Fern_ndez_2025}.
\end{enumerate}

\section{Computation of observables}
\label{sec:observables}

Details about the computation of the total energy in unconstrained as well as constrained dSCGF calculations can be found in~\cite{PRC}.
With total energies at hand, one- and two-neutron separation energies, as well as two-proton separation energies, can be computed via the differences
\begin{subequations}
\label{eq_sep_en}
\begin{align}
S_\text{1n}(N,Z) \equiv& |E^{(N,Z)}| - |E^{(N-1,Z)}| \; , \\
S_\text{2n}(N,Z) \equiv& |E^{(N,Z)}| - |E^{(N-2,Z)}| \; , \\
S_\text{2p}(N,Z) \equiv& |E^{(N,Z)}| - |E^{(N,Z-2)}| \; ,
\end{align}
\end{subequations}
respectively. Notice that one-nucleon separation energies can also be computed from the poles of the K\"all\'en-Lehmann representation of the single-particle propagator~\cite{PRC}.

The point-proton mean-square radius is extracted from the correlated one-body density matrix $\rho$ as
\begin{equation}
\langle r_p^2 \rangle = \frac{1}{Z} \sum_{\alpha\beta}^\text{protons} (r^2)_{\alpha\beta} \; \rho_{\beta\alpha}
\end{equation}
and converted into the charge mean-square radius through the standard formula
\begin{equation}
\langle r_{\rm ch}^2 \rangle = \langle r_p^2 \rangle + \langle R_p^2 \rangle + \frac{N}{Z}\langle R_n^2 \rangle + \frac{3\hbar^2}{4 m_p^2 c^2} + \langle r^2_{\rm so} \rangle \, ,
\end{equation}
where $\langle R_n^2\rangle=-0.1149$\,fm$^2$ and $\langle R_p^2\rangle=0.7079$\,fm$^2$ are respectively the neutron and proton mean-square charge radii, the second-to-last term represents the Darwin-Foldy correction, and $\langle r^2_{\rm so} \rangle$ is a spin-orbit contribution included through a spherical mean-field approximation~\cite{Horowitz12}. Two-body center-of-mass corrections are included following Ref.~\cite{Cipollone15}.

While the charge density distribution can be typically computed from the one-body density matrix, the symmetry-breaking character of the latter prevents a simple interpretation of the result and complicates the comparison with experiment.
Therefore, a spherical counterpart of $\rho$ is extracted from the deformed one by means of symmetry-adapted natural orbitals~\cite{Fasano22}.

Eventually, the charge density distribution is obtained from this symmetry-restricted density matrix following the procedure detailed in Ref.~\cite{Duguet17b}.

\section{Geometric origin of the $\beta_4(\beta_2)$ relation}

For a sharp, uniform-density surface with only axial quadrupole deformation,
\begin{equation}
R(\theta) = R_0\big[1+\beta_2 Y_{20}(\theta)\big] \, ,
\end{equation}
the multipole moment of order $\lambda$ is obtained by integrating the uniform density $\rho_0$ over the deformed volume,
\begin{equation}
\begin{split}
Q_\lambda &= \rho_0\int d\Omega\, Y_{\lambda 0}(\theta)\int_0^{R(\theta)} r^{\lambda+2}\,dr \\
&= \frac{\rho_0}{\lambda+3}\int d\Omega\, Y_{\lambda 0}(\theta)\, R(\theta)^{\lambda+3} \,.
\end{split}
\label{eq:Qlambda}
\end{equation}
Because $R(\theta)$ only contains the $Y_{20}$ harmonic, the powers $Y_{20}^n$ appearing in the binomial expansion of $R^{\lambda+3}$ decompose, by the triangle rule and parity selection of the Gaunt coefficients, only onto even multipoles $L\le 2n$. As the lowest two multipoles, $L=0,2$, are absorbed into volume conservation and the definition of $\beta_2$ itself, $\lambda=4$ is the first multipole genuinely generated by a pure quadrupole surface.

Setting $\lambda=4$ in Eq.~\eqref{eq:Qlambda} and expanding $(1+\beta_2 Y_{20})^{7}$ binomially, one obtains
\begin{equation}
\begin{split}
Q_4 = \frac{\rho_0 R_0^7}{7}\int d\Omega\, Y_{40}\Big[&1+7\beta_2 Y_{20}+21\beta_2^2 Y_{20}^2 \\
&+35\beta_2^3 Y_{20}^3+\mathcal{O}(\beta_2^4)\Big] \,.
\end{split}
\end{equation}
The first two terms vanish by orthogonality of $Y_{40}$ with $Y_{00}$ and $Y_{20}$. The next two terms  are non zero and are evaluated using the standard Gaunt integral
\begin{equation}
\begin{split}
\int d\Omega\, Y_{l_1 0}Y_{l_2 0}Y_{l_3 0}
=&\sqrt{\frac{(2l_1{+}1)(2l_2{+}1)(2l_3{+}1)}{4\pi}} \\
&\times\begin{pmatrix} l_1 & l_2 & l_3\\ 0&0&0\end{pmatrix}^{\!2},
\end{split}
\end{equation}
together with the $3j$-coefficients $\begin{pmatrix}4&2&2\\0&0&0\end{pmatrix}^{\!2}=\dfrac{2}{35}$ and $\begin{pmatrix}2&4&4\\0&0&0\end{pmatrix}^{\!2}=\dfrac{20}{693}$, which give
\begin{equation}
\int Y_{40}Y_{20}^2\,d\Omega = \frac{3}{7\sqrt{\pi}}\,,
\end{equation}
\begin{equation}
\int Y_{20}Y_{40}^2\,d\Omega= \frac{10\sqrt5}{77\sqrt{\pi}}\,.
\end{equation}
The cubic term additionally requires re-expanding $Y_{20}^2=\sum_{L=0,2,4}c_L\,Y_{L0}$, with $c_2=\sqrt5/(7\sqrt\pi)$ and $c_4=3/(7\sqrt\pi)$ obtained from the same Gaunt formula, so that
\begin{equation}
\begin{split}
\int Y_{40}Y_{20}^3\,d\Omega &= c_2\!\int Y_{40}Y_{20}^2\,d\Omega + c_4\!\int Y_{40}Y_{20}Y_{40}\,d\Omega \\
&= \frac{9\sqrt5}{77\pi}\,.
\end{split}
\end{equation}

Comparing to the linear-order relation defining $\beta_4$, $Q_4|_{\beta_4\text{-only}}=\rho_0 R_0^7\beta_4$, the hexadecapole moment generated by a pure quadrupole surface is
\begin{equation}
\begin{split}
\beta_4(\beta_2) &\approx \frac{1}{7}\Big[21\,\beta_2^2\!\int Y_{40}Y_{20}^2\,d\Omega + 35\,\beta_2^3\!\int Y_{40}Y_{20}^3\,d\Omega\Big] \\
&= \frac{9}{7\sqrt{\pi}}\,\beta_2^2 + \frac{45\sqrt5}{77\pi}\,\beta_2^3 + \mathcal{O}(\beta_2^4) \,.
\end{split}
\label{eq:beta4final}
\end{equation}
The leading coefficient, $9/(7\sqrt\pi)\approx0.725$, reproduces the quadratic term reported in Ref.~\cite{Xu2024}; the cubic term completes the expansion at this order.

\end{document}